\documentstyle[aps,twocolumn]{revtex}

\begin{document}
\draft
\title{Interplay between Coulomb Blockade and Resonant Tunneling 
studied by the Keldysh Green's Function Method}
\author{Tetsufumi Tanamoto}
\address{Research and Development Center, 
Toshiba Corporation, 
Saiwai-ku, Kawasaki 210, Japan}
\author{Masahito Ueda}
\address{Department of Physical Electronics, Hiroshima University,
Higashi-Hiroshima 739, Japan}
\date{\today}
\maketitle
\begin{abstract}
A theory of tunneling through a quantum dot is presented which
enables us to study combined effects of Coulomb blockade 
and discrete energy spectrum of the dot.  
The expression of tunneling current is derived from the 
Keldysh Green's function method, and is shown to automatically 
satisfy the conservation at DC current of both junctions. 
\end{abstract}

\medskip


\narrowtext

Over the past decade, the Coulomb blockade of tunneling has 
attracted growing interest mainly because of its possible 
applications in single-electron electronics\cite{Likarev}.
In view of ongoing advances in microfabrication 
techniques of semiconductor devices whose discrete 
energy-level structures are prominent, 
it seems urgent to develop a general theory 
of tunneling which 
allows us to analyze combined effects of Coulomb blockade and
discrete energy-level spectrum of the quantum dot.
While a number of articles study the combined effects
\cite{Groshev,Beenakker,Mier,Averin2,Iman}, 
in a usual perturbative treatment, 
the current conservation has been satisfied by adjusting 
a fitting parameter such as a electrochemical potential 
of the island. 
By generalizing the Keldysh equation developed by 
Caroli {\it et al.}\cite{Caroli},
we obtain an expression for the $I$-$V$
characteristics such that the  
currents at the left and right
junctions satisfy the conservation of DC current
\cite{Hershfield}. 
We also find that the half-width of the resonant peak( 
written by $\Gamma_{nn'}$ below ) 
is strongly modulated by the Coulomb interaction. 

The system consists of two tunnel junctions with capacitances $C_L$ 
and $C_R$, and a gate capacitor $C_g$ connected to the central island.
The Hamiltonian consists of the electronic part $\hat{H}_{\rm E}$, 
the transfer part $\hat{H}_{\rm T}$, and the part $\hat{H}_{\rm EM}$
that describes the electromagnetic (EM) environment surrounding 
the junctions.  
The electronic part consists of 
electrode and island parts, 
\begin{eqnarray}
\hat{H}_{\rm E} =  \! \! \! \!
\sum_{{\bf k}, \alpha \in L,R}  \! \! \! \! 
E_{{\bf k}\alpha} \hat{c}_{{\bf k}\alpha}^\dagger 
\hat{c}_{{\bf k}\alpha} \! +\!
\sum_{mn}[ E_{m}\delta_{mn} \hat{d}_{m}^\dagger \hat{d}_{m} 
+V^{\rm imp }_{mn} \hat{d}_{m}^\dagger \hat{d}_{n}], 
\nonumber
\end{eqnarray}
where $\alpha$ represents a set of parameters that, together 
with wave vectors ${\bf k}$, completely specify the 
electronic state of the left (L) or right (R) electrode, 
and $m,n$ specify the energy levels of the central island.
The symbol $\delta_{mn}$ denotes the Kronecker's delta and
$V^{\rm imp }_{mn}$ describe matrix elements of impurity 
scatterings in the island.
The transfer part is described by
\begin{eqnarray}
\hat{H}_{\rm T} = \! \! \sum_{n{\bf k} \alpha \in L,R} 
\! \! [ V_{n {\bf k}\alpha} (t)
\hat{c}_{{\bf k}\alpha}^\dagger e^{i\hat{\phi}_{\alpha}} \hat{d}_{n}  
+ \mbox{h.c.} ],
\nonumber
\end{eqnarray}
where $\hat{\phi}_{\alpha} \ (\alpha=L,R)$ are the phase operators 
which are canonically conjugate to the charge operators 
$\hat{Q}_\alpha$ 
on the junctions, satisfying the commutation relations 
$[\hat{\phi}_\alpha ,\hat{Q}_\beta]= ie \delta_{\alpha \beta}$
\cite{Devoret}. 
The EM part is described by
\begin{eqnarray}
& & \hat{H}_{\rm EM} =\! \frac{\hat{Q}_L^2}{2C_L} 
+\!\frac{\hat{Q}_R^2}{2C_R} \!+\!\frac{\hat{Q}_g^2}{2C_g} 
-\hat{Q}_L V\! -\hat{Q}_g V_g 
+ H(\{ \hat{\phi}_\alpha \}) 
\nonumber \\
& & =\! \frac{(\hat{Q} \!-\! C_0 \! V)^2}{2C_x} \! + \! 
\frac{(\hat{q}\!-\! C_g \! V_g \!- \! C_L \! V)^2}{2C_\Sigma}
\! +\! H(\{ \hat{\phi}_\alpha \})
+\! {\rm const.}, 
\label{EM}
\end{eqnarray}
where $C_0 \! \! \equiv \! \! C_L C_R /(C_L \! +\! C_R)$, 
$C_\Sigma \! \! \equiv \! \! C_L \! \! + \! \! C_R \! \!+ \! \! C_g$,
$C_x \equiv C_\Sigma C_0^2/[C_L (C_R\!+\!C_g)]$,
and $ H(\{ \hat{\phi}_\alpha \})$ describes the phase part of the
Hamiltonian whose concrete form depends on the external circuit. 
In the second line of Eq. (\ref{EM}), new
operators $\hat{q}$ and $\hat{Q}$ are introduced such that
\begin{eqnarray}
\frac{\hat{Q}_L}{C_L} &=& \frac{C_R+C_g}{C_\Sigma} 
\frac{\hat{Q}}{C_0}
+ \frac{\hat{q}-C_g V_g}{C_\Sigma}, \nonumber\\
\frac{\hat{Q}_R}{C_R} 
&=& \frac{C_L}{C_\Sigma} \frac{\hat{Q}}{C_0}
- \frac{\hat{q}-C_g V_g}{C_\Sigma}, \nonumber\\
\frac{\hat{Q}_g}{C_g} 
&=&-\frac{C_L}{C_\Sigma} \frac{\hat{Q}}{C_0}
 +\frac{\hat{q}-C_g V_g}{C_\Sigma}+ V_g.\nonumber
\end{eqnarray}
The commutation relations for the new operators are given by
$[\hat{\psi}, \hat{q}]=ie$ and 
$[\hat{\varphi}, \hat{Q}]=ie$, where
\begin{equation} 
\hat{\phi}_L= \hat{\psi}+\kappa_L \hat{\varphi},
\ \ 
\hat{\phi}_R=-\hat{\psi}+\kappa_R \hat{\varphi}
\label{eqn:phi}
\end{equation}
with $\kappa_L\equiv C_R/(C_L+C_R)$ and
$\kappa_R \equiv C_L/(C_L+C_R)$.

The current at junction $\alpha$ (=L or R) 
is given by\cite{Jauho}
\begin{equation}
J_\alpha (t) 
=(-1)^{\beta} \frac{2e}{\hbar} \mbox{Re} \left\{ \sum_{{\bf k}n} 
V_{n {\bf k}\alpha} (t) \tilde{G}^{<}_{n{\bf k} \alpha} (t,t) \right\},
\label{eqn:J0}
\end{equation}
where $\beta$=0 for the left junction and $\beta$=1 
for the right junction 
and $\tilde{G}_{n{\bf k}\alpha}^< (t,t')  \equiv  
i\langle \hat{c}_{{\bf k}\alpha}^\dagger (t) 
e^{-i\hat{\phi}_\alpha (t) }
\hat{d}_{n} (t') \rangle $
is an analytic continuation 
of the contour-ordered Green's function 
$\tilde{G}^{<}_{n{\bf k} \alpha} (\tau, \tau')$ which is defined
in the interaction representation by
\begin{eqnarray}
\tilde{G}_{n{\bf k}\alpha} (\tau, \tau')  
&\equiv&  i\langle T_C \{ 
\hat{c}_{{\bf k}\alpha}^\dagger (\tau') 
e^{-i\hat{\phi}_\alpha (\tau') } 
\hat{d}_{n} (\tau) 
\nonumber \\
&\times & \! \exp \left(-\frac{i}{\hbar} \! \! \int_C \! 
\hat{H}_{\rm T} (\tau_1) d \tau_1 \right) 
\} \rangle, 
\label{eqn:ggtau}
\end{eqnarray}
where $T_C$ is the contour-ordering operator. 
We assume that electrons in the left and right electrodes
are noninteracting. 
Then the only nonvanishing terms in Eq.\ (\ref{eqn:ggtau}) are 
those in which $\hat{c}_{{\bf k}\alpha}^\dagger (\tau')$ 
is contracted with 
$\hat{c}_{{\bf k}\alpha} (\tau_1)$ in the exponential term. 
\begin{eqnarray}
\lefteqn{\tilde{G}_{n{\bf k}\alpha} (\tau, \tau')= \!  
\int_C \frac{d \tau_2}{\hbar} \! \sum_{m} \!  
V_{m{\bf k}\alpha}^* (\tau_2)\langle T_C \{ \!
\hat{c}^\dagger_{{\bf k}\alpha} (\tau)
\hat{c}_{{\bf k}\alpha}(\tau_2)  \} \rangle }
\nonumber \\
& & \times \langle T_C \{ {\rm e}^{\!-\!i\hat{\phi}_\alpha (\! \tau \! )} 
{\rm e}^{\! i\hat{\phi}_\alpha (\! \tau_2 \! )}
\hat{d}_n (\tau' \!) \hat{d}^\dagger_m (\tau_2 \!) 
\exp \! \! \left(\! -\! \frac{i}{\hbar} \! \! \int_C \! 
\hat{H}_{\rm T} (\tau_1) d \tau_1 \! \right) \} \rangle \label{apx}. 
\end{eqnarray}
To proceed with calculation, we decouple the term $\langle 
T_C \{ {\rm e}^{\!-\!i\hat{\phi}_\alpha (\! \tau \! )} 
{\rm e}^{i\hat{\phi}_\alpha (\! \tau_2 \! )}
\} \rangle$ on the right hand side of Eq. (\ref{apx}). 
In spite of this decoupling, the conservation of DC current is 
 satisfied as will be below. 
We then obtain, after analytic continuation to real time, 
\begin{eqnarray}
\tilde{G}^{<}_{n{\bf k}\alpha} (t,t') 
&=& \sum_m \int_{-\infty}^{\infty} \frac{dt_1}{\hbar}
 V^*_{n{\bf k}\alpha} (t_1) 
[ G_{nm}^r (t,t_1) \tilde{g}_{{\bf k}\alpha}^< (t_1, t')  \nonumber \\
& & + G_{nm}^< (t,t_1) \tilde{g}_{{\bf k}\alpha}^a (t_1, t') ],
\label{eqn:Gd}
\end{eqnarray}
where $G^{<}_{nm}(t,t') \equiv i \langle 
\hat{d}_m^\dagger (t') \hat{d}_n (t) \rangle$ 
is the Green's function of the central island,  
and $\tilde{g}^<_{{\bf k}\alpha} (t_1,t_2 )$  
and 
$\tilde{g}^a_{{\bf k}\alpha} (t_1,t_2 )$ 
are given in terms of 
$g^{<}_{{\bf k}\alpha}(t_1, t_2 ) \equiv i \langle 
\hat{c}_{{\bf k} \alpha}^\dagger (t_2) 
\hat{c}_{{\bf k} \alpha} (t_1) \rangle$ 
and $ P_\alpha^{<} (t_1, t_2) \equiv \langle 
e^{i \hat{\phi}_\alpha (t_2)} 
e^{-i \hat{\phi}_\alpha (t_1)} \rangle$
as
$\tilde{g}^<_{{\bf k} \alpha} (t_1,t_2 ) 
=  g^{<}_{{\bf k} \alpha} (t_1,t_2) 
P^{<}_{\alpha} (t_1,t_2)  $ and 
$ \tilde{g}^a_{{\bf k} \alpha} (t_1,t_2 )   
= \theta (t_2-t_1)  [g^{<}_{{\bf k} \alpha} (t_1,t_2) 
P^{<}_{\alpha} (t_1,t_2) - \ 
g^{>}_{{\bf k} \alpha} (t_1,t_2) 
P^{>}_{\alpha} (t_1,t_2)]$.
Here 
$g^{>}_{{\bf k} \alpha} (t_1,t_2) \equiv 
-i \langle 
\hat{c}_{{\bf k} \alpha} (t_1) 
\hat{c}_{{\bf k} \alpha}^\dagger (t_2) \rangle$
and 
$P^{>}_{\alpha} (t_1,t_2) \equiv 
\langle 
e^{-i \hat{\phi}_\alpha (t_1)} 
e^{i \hat{\phi}_\alpha (t_2)} \rangle$.
Substituting Eq. (\ref{eqn:Gd}) into Eq. (\ref{eqn:J0}), 
we obtain 
\begin{eqnarray}
\lefteqn{ J_\alpha (t) \!  =(-1)^{\beta} \frac{e}{\hbar^2} \! 
\sum_{{\bf k} mn} \! V_{n{\bf k} \alpha}(t) \!
\int_{-\infty}^{\infty}  \! \! \! \! \!
dt' V^*_{m{\bf k} \alpha} (t') \{ G^>_{nm} (t,t') 
} \nonumber \\  & & \times  
g^<_{{\bf k} \alpha} (t',t) P^<_{\alpha} (t',t)  
-  G^<_{nm} (t,t') g^>_{{\bf k} \alpha} (t',t) 
P^>_{\alpha} (t',t) \} ,
\label{eqn:jstrt}  
\end{eqnarray}
where $G^>_{nm}(t,t')\equiv -i\langle 
\hat{d}_n (t) \hat{d}_m^\dagger (t')
\rangle$.
By introducing a noninteracting self-energy, 
$\Sigma_0^{><}$, defined by $G_0^{><} = G^r_0 \Sigma_0^{><} G^a_0$, 
and using the Dyson equations, $(1+G^r \Sigma^r)G^r_0=G^r$ 
and $G^a_0 (1+\Sigma^a G^a)=G^a$,
we may cast the Keldysh equation
$G^{><} =(1+G^r \Sigma^r)G_0^{><} (1+\Sigma^a G^a) 
+ G^r \Sigma^{><} G^a$ into the following form  \cite{Sarker}:
\begin{eqnarray}
G_{nm}^{><} (t,t') 
&=& \! \!  \sum_{n_1,n_2}
\! \int \! dt_1 dt_2 G^r_{nn_1} (t,t_1) 
\Sigma^{><}_{{\rm tot} \ n_1n_2} (t_1,t_2) \nonumber \\
&\times &G^a_{n_2 m} (t_2,t') ,  
\label{Keldysh eq}
\end{eqnarray}
where
\begin{eqnarray}
\Sigma^{><}_{{\rm tot} \ n_1n_2} (t_1,t_2)
&=& \Sigma^{><}_{0 \ n_1n_2} (t_1,t_2)
+ \Sigma^{><}_{{\rm T} \ n_1n_2} (t_1,t_2) \nonumber \\
&+& \Sigma^{><}_{{\rm s} \ n_1n_2} (t_1,t_2).
\label{eqn:eta0}
\end{eqnarray}
The first term on the right-hand side describes the self-energy
of the island in the absence of disorder, interaction, and tunneling. 
Since it corresponds to the free part, it is infinitesimal:
$\Sigma^{><}_{0n_1 n_2} (\epsilon)= 2i \delta_{n_1 n_2} \eta 
[f_0 (\epsilon)-1/2 \mp 1/2]$, 
where $\eta$ is a positive infinitesimal, $>(<)$ refers to the 
minus (plus) sign,
$f_0(\epsilon) = (\exp \beta (\epsilon-V_d) +1 )^{-1}$, 
and  $V_d \equiv (C_L V  + C_g V_g)/C_{\Sigma} $.
The second term describes a self-energy due to tunneling:
\begin{eqnarray}
\lefteqn{ \Sigma^{><}_{{\rm T} n_1n_2} (t_1,t_2) }
\nonumber \\ &=&
\sum_{{\bf k} \alpha} 
\frac{V_{n_1 {\bf k} \alpha}^*(t_1) 
V_{n_2 {\bf k} \alpha}(t_2) }{\hbar^2}
g_{{\bf k} \alpha}^{><} (t_1,t_2) P_\alpha^{><} (t_1,t_2),
\label{eqn:sigma}
\end{eqnarray}
and the third term describes effects of scattering:
$\Sigma^{><}_{{\rm s} \ n_1n_2} (\epsilon)= 
i \delta_{n_1 n_2} [f_d (\epsilon)-1/2 \mp 1/2]/\tau_{\rm s}$, 
where
$\tau_{\rm s}$ is a scattering time in the island and 
$f_d(\epsilon)$ is a Fermi distribution function in 
the presence of the scattering. 
Because of the infinitesimal factor $\eta$,
the free part is  important only when the remaining parts 
are absent.

In the following we focus on the case in which
$\Gamma_{nn'{\bf k}\alpha} (t,t') \equiv (2\pi/\hbar)
V_{n{\bf k} \alpha}^* (t)V_{n'{\bf k} \alpha}(t')$ 
is a real function of $t-t'$.  
From Eq.\ (\ref{eqn:jstrt}), $J_L-J_R=e\int_{-\infty}^{\infty} 
dt' (G_{nm}^{>}(t,t') \Sigma_{{\rm T}mn}^{<}(t',t)
-  G_{nm}^{<}(t,t') \Sigma_{{\rm T}mn}^{>}(t',t))$. 
Because $G_{nm}^{><}(t,t')$ contains 
$\Sigma_{{\rm T}mn}^{><}(t,t')$
(Eq. (\ref{eqn:sigma})), 
the conservation of current through the two 
junctions is automatically satisfied, that is, $J_L=J_R$.
This is a great step forward, considering the fact that in a usual
perturbative treatment one has to adjust the electrochemical 
potential of the island so as to satisfy the current 
conservation.

The retarded and advanced Fourier-transformed 
Green's functions at the central island, 
$G^r_{nn'} (\epsilon)$ and $G^a_{nn'} (\epsilon)$, 
are derived from the Dyson equation:
\begin{eqnarray}
\lefteqn{\hbar [G^{r,a}_{nn'} (\epsilon)]^{-1} 
= \hbar [ g^{r,a}_n (\epsilon) ]^{-1} 
- \hbar \Sigma^{r,a}_{{\rm tot} \ nn'} (\epsilon) }
\nonumber \\ 
&=& \! \epsilon \! -\! (E_n \! +\! V_d) 
\! - \! \Lambda_{nn'} (\epsilon) \! 
\pm \!  \frac{i}{2} \left( 2\eta \! +\! \frac{\hbar}{\tau_s} 
\! + \! \Gamma_{nn'} (\epsilon) \right),
\end{eqnarray} 
where 
$\Gamma_{nn'} (\epsilon) \equiv 2 {\rm Im} 
\Sigma^{a}_{{\rm T} nn'} (\epsilon)$, 
and $2\pi \hbar \Sigma^{r,a}_{{\rm T} nn'} (\epsilon) \! \! 
= \! \! \sum_{{\bf k} \alpha } 
\Gamma_{nn'{\bf k} \alpha} (\epsilon) 
\tilde{g}_{{\bf k} \alpha}^{r,a} (\epsilon)$.  
The real part of the self-energy 
$\Lambda_{nn'} (\epsilon)$, 
which are due to scattering and tunneling, 
shifts energy levels
in the central island and we will regard this effect as included 
in our assumed one-body energy levels of the quantum dot. 
Here $\Gamma_{nn'}(\epsilon)$ broadens resonant peaks and 
is strongly modulated by the Coulomb interaction through 
the $P^{><}$ functions:
\begin{eqnarray}
\lefteqn{
\Gamma_{nn'} (\epsilon) = \sum_{{\bf k} \alpha } 
\! \int \! \frac{d \epsilon_1}{2\pi \hbar}  
\Gamma_{nn'{\bf k} \alpha} (\epsilon_1) 
[ (1-f_\alpha (E_{{\bf k} \alpha}) )}\nonumber \\
&\times & P^>_\alpha ( \epsilon-\epsilon_1 -E_{{\bf k} \alpha} ) 
+f_\alpha (E_{{\bf k} \alpha}) 
P^<_\alpha ( \epsilon-\epsilon_1 -E_{{\bf k} \alpha} ) ], 
\label{eqn:GLR}
\end{eqnarray}
where $f_L(\epsilon) \! \! =\! \! 1/ \! 
( e^{\beta(\epsilon  -\! E_F \! -\! eV)} \! 
\! +\! 1 )$ and $f_R(\epsilon) \! \! = \! \! 
1/ \! ( e^{\beta(\epsilon -E_F)} \!  \! +\!  1 ) $ with 
$E_F$ being the Fermi energy of the electrodes. 

We show here that the free part of the self-energy can be 
neglected even in the absence of scattering. 
Because $G_{{\rm free} \ n}^{><} \equiv G_n^r 
\Sigma_{0}^{><} G_n^a$ is given by 
\begin{equation}
G_{{\rm free} \ n}^{><} (\epsilon ) \! =\! 
\frac{2\eta (f_0(\epsilon) -\frac{1}{2} \mp \frac{1}{2} )}
{(\epsilon \! -\! E_n \! -\!  V_d \!+ \! 
\Lambda_n (\epsilon) )^2 
\! +\! ( \eta \! + \! \Gamma_n (\epsilon)/4)^2 }, 
\label{gfree}
\end{equation}
$G_{{\rm free} \ n}^{><} (\epsilon )$ remains nonvanishing
only when $\Gamma_n (\epsilon)=0$ as can be seen
from the relation, 
$ \lim_{\eta  \rightarrow +0} 
\eta/[(\epsilon-E_n-V_d+\Lambda_n (\epsilon))^2 + \eta^2] 
= \pi \delta (\epsilon -E_n-V_d+\Lambda_n (\epsilon))$. 
If $\Sigma_{n\alpha}^{><} (t)  
\equiv \sum_{{\bf k}\alpha} \Gamma_{n{\bf k} \alpha}(t) 
g_{{\bf k}\alpha}^{><} (t) P_{\alpha}^{><}(t)/2\pi$, 
Eq. (\ref{eqn:jstrt}) is cast into 
\begin{equation}
\! J_{\alpha} \! \! =\! \!(-1)^{\beta}   
\frac{e}{\hbar^2} \! \sum_{n} \! \! \int 
\! \! \frac{d \epsilon_1}{2 \pi} \{ 
 G^>_{n} (\epsilon_1) \Sigma^<_{n \alpha}(\epsilon_1) 
\! -\! 
G^<_{n} (\epsilon_1) \Sigma^>_{n \alpha}(\epsilon_1) \} .
\label{crnt1}
\end{equation}
Because $\Gamma_{n} (\epsilon) =\sum_\alpha (
\Sigma_{n \alpha}^> (\epsilon) -
\Sigma_{n \alpha}^< (\epsilon))=0$ means that 
$\Sigma_{n \alpha}^> (\epsilon) =
\Sigma_{n \alpha}^< (\epsilon)=0$
[note that each term in Eq. (\ref{eqn:GLR}) is non-negative], the 
current disappears when $\Gamma_n ( \epsilon )=0$. 
Thus the free part can be neglected 
in the expression of current. 
In a usual resonant tunneling problem, 
$\Gamma_n =\Gamma_L + \Gamma_R$ 
(width at half maximum: constant), 
so  $\eta$ in the denominator can be neglected 
anyway. 

With  $ A_{n_1 n_2}(\epsilon) 
\equiv i (G^r_{n_1 n_2}(\epsilon) -G^a_{n_1 n_2} (\epsilon))$, 
and $\hbar \Theta_{nm} (\epsilon) \equiv 
G^r_{nn_1} (\epsilon) G^a_{n_2m} (\epsilon) /  
 A_{n_1 n_2} (\epsilon) $,  
$J_L$ in the absence of scattering 
can be cast into the following form,  
\begin{eqnarray}
\lefteqn{J_L = -\frac{e}{\hbar} 
\int_{-\infty}^{\infty} \frac{d \epsilon_1}{2\pi\hbar}
\sum_{{\bf kk'}nmn_1 n_2} \! \! \! \! \! \! \! \! \! \!  
\frac{V_{n{\bf k}L}^* (\epsilon_1)V_{m{\bf k}L}(\epsilon_1) 
V_{n_1 {\bf k'} R}^* (\epsilon_1) 
V_{n_2 {\bf k'} R}(\epsilon_1)}{\hbar^2}  
A_{n_1 n_2} (\epsilon_1 ) \Theta_{nm} (\epsilon_1) }\nonumber  \\
& & \times \left\{ f_{L} (E_{{\bf k}L}) 
\left( f_{R}(E_{{\bf k}'R})-1 \right) 
P_L^< (\epsilon_1 -E_{{\bf k}L})  
P_R^> (\epsilon_1 -E_{{\bf k}'R}) 
- \left( f_{L} (E_{{\bf k}L} )-1 \right) f_{R}(E_{{\bf k}'R}) 
P_L^> (\epsilon_1 -E_{{\bf k}L}) 
P_R^< (\epsilon_1 -E_{{\bf k}'R}) \right\}.
\label{eqn:mr}
\end{eqnarray}
This is the central result of this paper. 
In the absence of the charging effects, 
{\it i.e.}, $P_{\alpha}^{><}  
(\epsilon-\epsilon')=\delta (\epsilon-\epsilon')$, 
$J_L$ is reduced to the familiar expression of resonant tunneling 
\cite{Weil}.
When $\Sigma^{r,a}_{mn} (\epsilon) =
\delta_{mn} \Sigma^{r,a}_n$,  
all Green's functions are diagonalized. 
This corresponds, {\it e.g.}, to a situation in which 
 energy levels in the island are mutually uncorrelated 
during the tunneling process.  
Then $A_{n_1 n_2} (\epsilon)$ reduces to 
\begin{equation} 
A_n (\epsilon)= \frac{\hbar \Gamma_n (\epsilon) }
{\left[ \epsilon-( E_n +V_d )  
- \Lambda_n (\epsilon) \right]^2 
+\left[ \Gamma_n (\epsilon) \right]^2 /4 }.  
\label{an}
\end{equation} 
and $\Theta_n (\epsilon)$ reduces to $\Gamma_n (\epsilon)^{-1}$. 

From Eq.\ (\ref{eqn:phi}), we see that 
$P_\alpha^>$ can be divided into two parts:
$P_\alpha^> (t_1,t_2)= \langle e^{-i\kappa_\alpha 
\hat{\varphi} (t_1)} 
e^{i \kappa_\alpha \hat{\varphi} (t_2)} \rangle$ $\times 
\langle e^{-i\zeta_\alpha \hat{\psi} (t_1)} 
e^{i \zeta_\alpha \hat{\psi}(t_2)} \rangle$
($\zeta_L =1$ and $\zeta_R=-1$),  
where the first factor describes influence of the electrodynamic 
environment (i.e. the external circuit) \cite{Devoret}, 
and the second factor describes quantum fluctuations of the island 
charge.
Using the relation 
$\langle m | e^{-i \gamma \hat{\psi} (t)} 
e^{i \gamma \hat{\psi} (0)} | m \rangle 
= e^{-i\frac{U}{\hbar} [1+ 2\gamma (m -\bar{n}_c) ]t}$, where
$\bar{n}_c \! \! \equiv \! \!
(C_L \! V \! \! + \! \! C_g  \! V_g)/e$,  
$\gamma$ is an arbitrary constant \cite{Higurashi}, and $| m\rangle $ 
is the charge eigenstate of the island with charge $em$,
we obtain
\begin{displaymath}
\langle  e^{-i \gamma \hat{\psi} (t)} 
e^{i \gamma \hat{\psi}(0)} 
\rangle  = \! \! \! \sum_{[m]}
\! \!  
\frac{e^{-\beta U (m-\bar{n}_c)^2 }}{c_u}
 e^{-i\frac{U}{\hbar} [1+ 2\gamma (m -\bar{n}_c) ]t}, 
\end{displaymath}
where $U \! \! \equiv \! \! e^2 \! /2C_\Sigma$ and 
$c_u  \! \! \equiv  \! \! 
\sum_{[m]} \! 
e^{-\beta U (m-\bar{n}_c)^2 } $, 
and [m] runs over $1,2,\cdots,2N$, where $N$ is the number of
doubly degenerate energy levels.
Fourier transforming this in the high-impedance limit gives
\begin{equation}
\!  P^{><}_\alpha (\epsilon) \! =  \! \! \!  
\sum_{[m]} \! \!
\frac{e^{-\beta U (m-\bar{n}_c)^2 } }{c_u}
2 \pi \hbar \delta \! 
\left( \epsilon \mp E^{><}_{m \alpha}\!
\pm \kappa_\alpha^2 E_{C_x} \!\right) \!,
\label{eqn:PLR}
\end{equation}
where $E^{><}_{m \alpha} \equiv 
U (1 \pm 2[m -\bar{n}_c] \zeta_\alpha ) 
$, and
$E_{C_x} \equiv e^2/2C_x$.

Before we discuss the main consequences of Eq. (\ref{eqn:mr}),
let us discuss the relationship of our theory to the 
^^ orthodox theory' of double junctions \cite{Amman}.  
In the orthodox theory, 
the tunneling rate is assumed to be small, 
and  tunneling processes at the left and right junctions 
are treated separately by assuming that electrons 
on the island are always at thermal equilibrium.
To ensure this, we assume that inelastic scattering in 
the island is large and that tunneling is weak so that
$2\pi/\hbar |V_{n{\bf k} \alpha}^* (\epsilon) 
V_{n'{\bf k} \alpha}^* (\epsilon)| \ll
\hbar/2 \tau_{\rm s}  \ll E_F $, where $E_F$ is the Fermi energy of 
the central electrode.
Equation (\ref{eqn:jstrt}) can then be shown to reduce to the 
corresponding formula in the orthodox theory. 
In fact, under these assumptions, we have
${\rm Im}[G^{r,a}_k (\epsilon)]^{-1} \sim \pm \hbar/2 \tau_{\rm s}$ 
and  
$G^{><}$ may be approximated as $G_0^{><}$
because  near $\epsilon \sim E_F$ we have
$\lim_{\hbar/\tau_{\rm s} \rightarrow +0} 
\hbar/(2\tau_{\rm s})/[(\epsilon-E_n-V_d)^2 
+ (\hbar/2\tau_{\rm s})^2] 
= \pi \delta (\epsilon -E_n-V_d)$.
Thus, with appropriate identification of parameters, our formulation reduces 

\vspace*{2cm}

to 
that of Ref.\ \cite{Amman}.
In the following discussions, we consider the case in which the 
self-energy is diagonal [{\it i.e.} we assume Eq.\ (\ref{an})] and 
$\Gamma_{nn'{\bf k} \alpha}(\epsilon)$ does not depend on 
$n,n'$ and ${\bf k}$, namely, $\Gamma_{nn'{\bf k} \alpha}(\epsilon)
=\Gamma_{\alpha}(\epsilon)$ 
and $C_g \rightarrow 0$ and $V_d =C_L/C_\Sigma V+V_g$.

Figures \ref{metallic1} and  \ref{metallic2} show 
$I$-$V$ characteristics and the $I$-$V_g$ characteristics, 
when the charging energy $U\equiv e^2/2C_\Sigma$ 
is much greater than the discrete energy-level spacing $\Delta E_n$. 
From the relation  
$\Gamma_{\alpha} (\epsilon)= 2 \pi D_k (\epsilon)
|V_{{\bf k}\alpha}|^2$,
resistances $R_\alpha (\alpha=L, R)$ can be evaluated 
to be $ R_\alpha /R_K = (\Gamma_{\alpha}(E_F) D_d (E_F))^{-1}$,
where $R_K$ is the resistance quantum 
$h/e^2$ =25.8k$\Omega$ and $D_d (E_F)$ is 
the DOS of the island  at the Fermi energy. 
The resistances are estimated to be 
$R_R /R_K \sim 40$(weak tunneling)  when $D_d (E_F) E_F \sim 1$. 

Figure \ref{resonant} shows the $I$-$V$ characteristics in
which 
$\Delta E_n$ =25meV larger than the elementary charging
energy $U$=0 or 10meV. 
Here we assume $\Gamma_\alpha (\epsilon) =\sqrt{\epsilon/E_F}
 \Gamma_{\alpha 0}$ 
for the energy dependence of the 
density of state (DOS) $D_k (\epsilon)$ 
of the electrodes on the 
$I$-$V$ characteristics. 
In the absence of the Coulomb interaction, i.e., $U=0$, 
equally-spaced resonant tunneling peaks are seen. 
The Coulomb interaction deforms this $I$-$V$ characteristic 
in a manner 
depending on the strength of the interaction and temperature. 
As the Coulomb interaction increases, the deformation of 
the peaks becomes more pronounced, 
and some peaks are suppressed by large Coulomb gaps. 
At $T$=100K with $U$=10meV, the resonant peaks are 
thermally blurred.
The dips at 0.2 eV and 0.4 eV appear  
where original resonant peaks overlap with Coulomb gaps. 
Our results shows that the complex $I$-$V$ characteristics appear 
due to combined effects of Coulomb 
blockade and resonant tunneling in the quantum dot. 

In conclusion, we have studied combined effects of
Coulomb blockade and resonant tunneling 
through a double-barrier system 
by using the Keldysh equation.  
A general expression of the $I$-$V$ 
characteristics is obtained which 
automatically satisfies the conservation of DC currents 
at both junctions. 
We  found that, in a resonant-tunneling-dominated
regime, some of resonant tunneling peaks 
are suppressed due to the charging effects. 

T. T. acknowledges M. Azuma and Y. Ikawa 
of Toshiba Corp. for support throughout this work, 
and  A. Kurobe of Toshiba Corp. 
and S. Iwabuchi of Nara Woman's Univ. 
for fruitful discussions.
M.U. acknowledges support by a Grant-in-Aid for Scientific
Research (Grant No. 08247105) by the Ministry of Education,
Science, Sports, and Culture of Japan, and by the Core 
Research for Evolutional Science and Technology (CREST) of
the Japan Science and Technology Corporation (JST).


\begin{figure}
\caption{$I$-$V$ characteristics in the continuum limit 
for two different gate voltages  $V_g$=0 and -0.5eV
and two temperatures $T$=4.2 and 30K, where 
$E_F$=0.2eV, $U$ =5meV, $\Gamma_L$=5meV, 
$\Gamma_R$=0.1 meV.}
\label{metallic1}
\end{figure}

\begin{figure}
\caption{Coulomb oscillations in the continuum limit 
for $\Gamma_L$=5meV, 
$\Gamma_R$=0.1meV
with varying values of the elementary charging energy 
$U\equiv e^2/2C_\Sigma$, 
where, $E_F$=0.2eV, $T$=4.2K, and $V$=0.01eV. }
\label{metallic2}
\end{figure}

\begin{figure}
\caption{Combined effects of the Coulomb blockade and 
resonant tunneling for $T=1$ and 100K, where 
$E_F$ =0.07eV ($T$=1K), $V_g$ =-0.05eV,  
$\Gamma_{L0}$=$6.0\times 10^{-4}$meV, and 
$\Gamma_{R0}$=$4.0\times 10^{-4}$meV. 
The curves at $T$=100 K refer to the right scale and those 
at 1 K refer to the left one.}
\label{resonant}
\end{figure}



\begin{references}

\bibitem{Likarev}
{\it Single Charge Tunneling}, edited by H. Grabert 
and M. Devoret, NATO ASI Ser. B, Vol. 294 (Plenum, 
New York, 1991).

\bibitem{Groshev}
A. Groshev: Phys. Rev. B {\bf 42} 5895 (1990).

\bibitem{Beenakker}
C. W. Beenakker: Phys. Rev. B {\bf 44} 1646 (1991).
\bibitem{Mier}
Y. Mier, N. S. Wingreen and P. A. Lee:  
Phys. Rev. Lett. {\bf 66}, 3048 (1991). 

\bibitem{Averin2}
D. V. Averin, A. N. Korotkov and K. K. Likharev: 
Phys. Rev. B {\bf 44}, 6199 (1991).

\bibitem{Iman}
H. T. Iman, V. V. Ponomarenko and D. V. Averin 
:Phys. Rev. B {\bf 50}, 18288 (1994).

\bibitem{Caroli}
C. Caroli, R. Combescot, P. Nozieres and D. Saint-James: 
J. Phys. C. {\bf 4}, 916 (1971).

\bibitem{Hershfield}
An approximate treatment of the Coulomb interaction 
could break the current conservation: see, for example, 
S. Hershfield, J. H. Davies and W. Wilkins, 
Phys. Rev. B {\bf 46}, 7047 (1992).

\bibitem{Devoret}
H. Grabert, G.-L. Ingold, M. H. Devoret, D. Est\`eve, H. Pothier 
and C. Urbina: Z. Phys. B {\bf 84}, 143 (1991).

\bibitem{Jauho}
A. P. Jauho, N. S. Wingreen and Y. Mier, 
Phys. Rev. B {\bf 50}, 5528 (1994); Y. Mier and N. S. Wingreen, 
Phys. Rev. Lett. {\bf 68} 2512 (1992). 

\bibitem{Sarker}
S. K. Sarker: Phys. Rev. {\bf B32} 743 (1985).

\bibitem{Weil}
T. Weil and B. Vinter, 
Appl. Phys. Lett. {\bf 50}, 1281 (1987); 
M. Jonson and A. Grincwaig, {\it ibid} {\bf 51}, 1729 (1987)

\bibitem{Higurashi}
H. Higurashi, S. Iwabuchi and Y. Nagaoka, 
Phys. Rev. B. {\bf 51}, 2387 (1995).

\bibitem{Amman}
M. Amman, R. Wilkins, E. Ben-Jacob, P. D. Maker and 
R. C. Jaklevic: Phys. Rev. {\bf B 43} 1146 (1991)

\end{references}
\end{document}